\documentclass[chi_draft]{sigchi}

\CopyrightYear{2018}
\setcopyright{acmlicensed}
\doi{http://dx.doi.org/10.1145/3173574.3174219}
\isbn{978-1-4503-5620-6/18/04}
\conferenceinfo{CHI'18,}{April 21--26, 2018, Montreal, Canada}
\acmPrice{\$15.00}

\usepackage{balance}       
\usepackage{graphics}      
\usepackage[T1]{fontenc}   
\usepackage{txfonts}
\usepackage{mathptmx}
\usepackage[pdflang={en-US},pdftex]{hyperref}
\usepackage{color}
\usepackage{booktabs}
\usepackage{textcomp}

\usepackage{microtype}        
\usepackage{ccicons}          
\usepackage[utf8]{inputenc} 

\usepackage{subfig}
\usepackage[font={small,bf},skip=2pt]{caption}
\def\plaintitle{Breeze: Sharing Biofeedback Through\\ Wearable Technologies}
\def\plainauthor{Jérémy Frey, May Grabli, Ronit Slyper, Jessica R. Cauchard}
\def\plainkeywords{Affective Computing; Physiological Computing; Breathing; Wearables; Signal Processing}

\makeatletter
\def\url@leostyle{%
  \@ifundefined{selectfont}{
    \def\UrlFont{\sf}
  }{
    \def\UrlFont{\small\bf\ttfamily}
  }}
\makeatother
\def\pprw{8.5in}
\def\pprh{11in}

\definecolor{linkColor}{RGB}{6,125,233}
\hypersetup{%
  pdftitle={\plaintitle},
  pdfauthor={\plainauthor},
  pdfkeywords={\plainkeywords},
  pdfdisplaydoctitle=true, 
  bookmarksnumbered,
  pdfstartview={FitH},
  colorlinks,
  citecolor=black,
  filecolor=black,
  linkcolor=black,
  urlcolor=linkColor,
  breaklinks=true,
  hypertexnames=false
}

\def\sharedaffiliation{\end{tabular}   \begin{tabular}{c}}

\begin{document}

\title{\plaintitle}

\numberofauthors{4}
\author{%
  \alignauthor Jérémy Frey$^{*,\ddag}$\\
    \email{jfrey@ullo.fr}\\
  \alignauthor May Grabli$^{*}$\\
      \email{may.grabli@post.idc.ac.il}\\
  \alignauthor Ronit Slyper$^{*}$\\
    \email{rys@cs.cmu.edu}\\
  \alignauthor Jessica~R.~Cauchard$^{*}$\\
    \email{jcauchard@acm.org}\\
      \sharedaffiliation
      \affaddr{$^{*}$Ubiquitous Computing Lab, Interdisciplinary Center (IDC) Herzliya, Herzliya, Israel}\\
      \affaddr{$^{\ddag}$Ullo, La Rochelle, France} \vspace{0.25cm}\\}
\maketitle

\begin{abstract}
Digitally presenting physiological signals as biofeedback to users raises awareness of both body and mind. This paper describes the effectiveness of conveying a physiological signal often overlooked for communication: breathing. 
We present the design and development of digital breathing patterns and their evaluation along three output modalities: visual, audio, and haptic. We also present Breeze, a wearable pendant placed around the neck that measures breathing and sends biofeedback in real-time.
We evaluated how the breathing patterns were interpreted in a fixed environment and gathered qualitative data on the wearable device’s design. We found that participants intentionally modified their own breathing to match the biofeedback, as a technique for understanding the underlying emotion. 
Our results describe how the features of the breathing patterns and the feedback modalities influenced participants' perception. We include guidelines and suggested use cases, such as Breeze being used by loved ones to increase connectedness and empathy. 

\end{abstract}

\category{H.5.m.}{Information Interfaces and Presentation
  (e.g. HCI)}{Miscellaneous}
  

\section{Introduction}
\begin{figure}
\includegraphics[width=1\hsize]{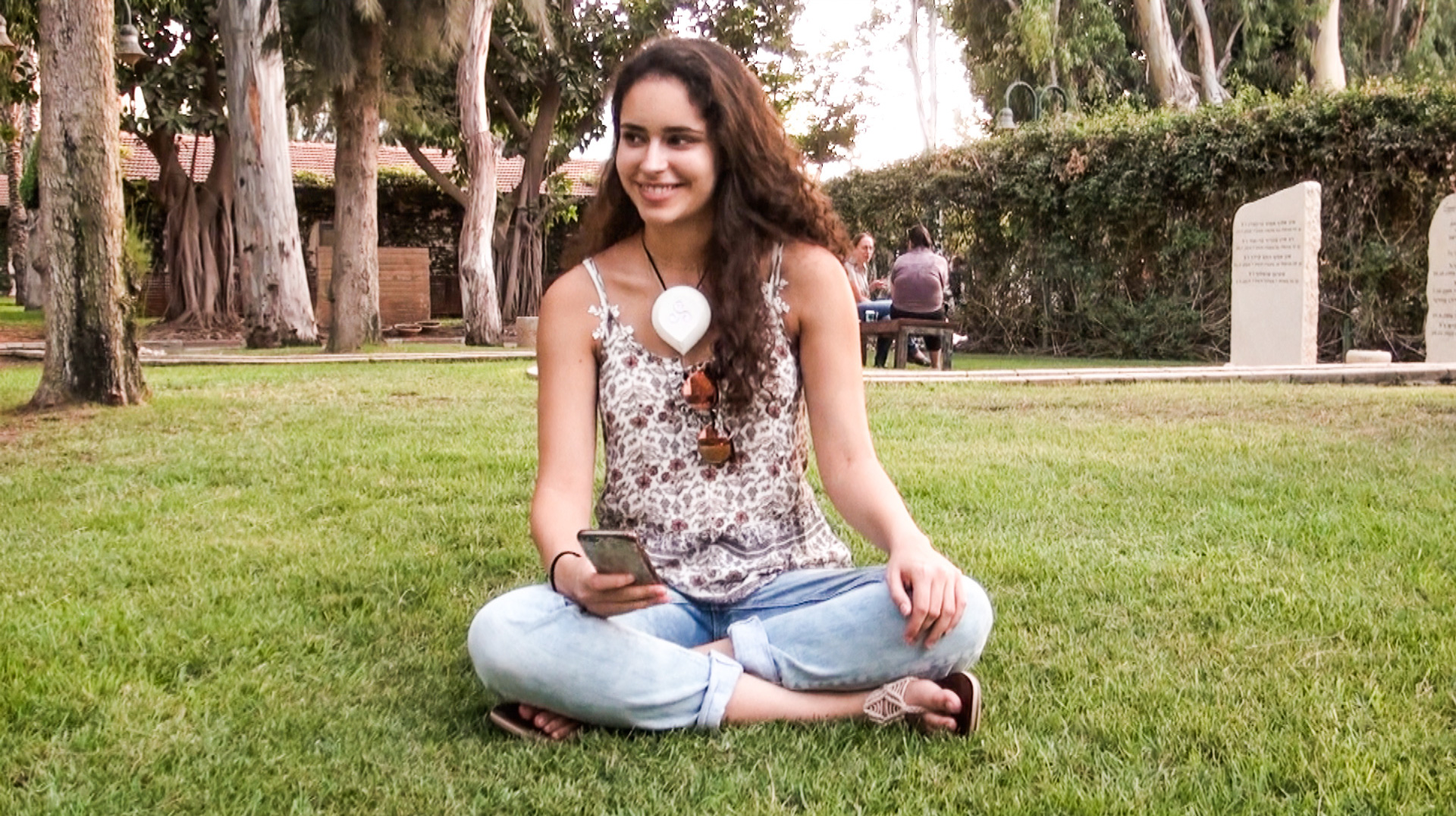}
\caption{User with Breeze, a pendant for shared breathing biofeedback.}\label{fig:teaser}
\end{figure}
This research investigates how one perceives emotions through breathing patterns, and how a wearable device can serve as both a sensor and as a way to convey biofeedback. 

Biofeedback is meant to make explicit a physiological signal, in such a way that it becomes more noticeable. The feedback shifts people's attention to their internal processes, raising awareness of body and mind. The display of biofeedback triggers a feedback loop that can be used for self-regulation \cite{Costa:2016:EmotionCheck,Moran2016}, even with high-level signals such as neuronal activity \cite{Sitaram2016}. Since contemplative practice and interoception -- one's ability to sense one's inner states -- are closely related to well-being, biofeedback may improve health \cite{Farb2015}.

We advocate for breathing as a source of biofeedback that can convey a variety of information, both for internal awareness and for mediating between people.

Breathing is tightly intertwined with other rhythms of the body, maintaining a close relationship with neural correlates \cite{Homma2008}. From an evolutionary perspective, breathing is closely linked to emotions: it plays an important role in ``fight or flight'' responses \cite{VanDiest2001} and through smell helps to sense dangers in the environment \cite{Masaoka2014}. 

Breathing is also affected by our interactions with others, as when taking turns in a discussion \cite{Rochet-Capellan2014}. More broadly, it appears that in some situations physiological mechanisms support interpersonal coordination. Such synchrony plays an important role in social behaviors and empathy \cite{Keller2014}, allowing biofeedback displays to be used for social interactions \cite{Chanel2015}. 

Signals such as heart rate and electrodermal activity (EDA), which measures perspiration, can also serve as proxies for social interactions \cite{Merrill2017}. Breathing presents an advantage over such signals because it can easily be modulated. In the field of human-computer interaction (HCI), breathing and other physiological signals have been used mainly as input \cite{Marshall2011,Nacke2011a}. Instead, we employ these technologies to support \emph{human-human communication} as advocated by \cite{Brueckner2014}.

We envision a future where interfaces can ``connect'' people at a distance, enforcing a bond between close ones through shared wearable biofeedback. While trying to develop such an interface, we identified gaps in the literature. Several theoretical and technical challenges remain to be addressed, such as: What is the best way to digitally encode breathing? How do people perceive breathing patterns across multiple modalities? Are people affected by perceiving another person's breathing? Which non-invasive sensors can be used for such a task? Is biofeedback perceived differently across multiple modalities? 

These gaps motivated the work as a first stepping stone advancing fundamental knowledge into the use of multiple modalities for biofeedback. This paper discusses the related work; the design and development of the Breeze prototype; a fourteen-participant indoor study evaluating digitally encoded breathing patterns and how users react to them; and the results and conclusions of the study. 

The paper's contributions are as follows:
\begin{enumerate}
\item Breeze: a wearable pendant that monitors a person's breathing and communicates it to a connected companion's pendant (Figure \ref{fig:teaser}). The pendant can be used for both input and multi-modal output.
\item An understanding of how people perceive and interpret breathing biofeedback across three modalities: visual, audio, and haptic.
\item A ``lexicon of breathing'' that exposes the relationships between breathing features and perceived emotions.
\item Guidelines toward shared breathing biofeedback. 
\end{enumerate}

\section{Related work}
Breathing as biofeedback has been studied from several angles. Gervais et al. use tangible avatars to display the biofeedback of two participants tasked with synchronizing their breathing to foster calmness \cite{Gervais2016}. In \cite{Ghandeharioun2017}, the authors used slowed respiratory feedback to reduce stress in computer users. More recently, breathing was shown to affect the emotional expressiveness of robots \cite{Bucci2017}.


In the next subsections, we first investigate which breathing features elicit emotion. We then describe technologies that measure breathing. Finally, we cover various modalities for displaying biofeedback.

\subsection{The Language of Breathing}
While the modulation of breathing is associated with changes in cognitive processes such as workload \cite{Rennert2013}, most previous work focuses on the link between breathing and emotions.

Notably, Boiten et al. conducted an extensive review of existing literature \cite{Boiten1994} and studied breathing responses during film scenes \cite{Boiten1998}. Among the traits described:

\begin{itemize}
\item \textbf{Inspiratory length} A shorter amount of time spent inhaling was correlated with pleasantness, whereas a longer inhalation was induced by tension and excitement.
\item \textbf{Holding the breath} Amusement and disgust elicited a longer pause after inspiration, whereas rest and relaxation were linked to a longer pause after expiration.
\item \textbf{Pace} Increase in excitement led to more rapid breathing; calm led to slower breathing.
\item \textbf{Deep/Shallow} Slow and deep breaths were linked to excitement and positive emotions, while slow and shallow breaths were linked to calm. Regular and shallow breathing was linked to concentrating. Deeper breathing was also linked to negative emotions.
\item \textbf{Irregularities} Irregular breathing could be caused by excitement. Holding the breath in the middle of the cycle was a sign of surprise.
\end{itemize}
We use these insights as guidelines in our user study; however, other factors must be taken into account: 1) many interactions (e.g., between amplitude and pace) obfuscate the relationship between emotion and breathing; 2) these effects were recently shown to change depending on age and gender \cite{Gomez2016}; and 3) measuring tools are imperfect. For example, while the ratio between thoracic and abdominal respiration varies with the emotion, few techniques account for the discrepancy.

\subsection{Measuring Breathing}
Breathing is measured using airflow, with a sensor put next to the nose or the mouth, or by detecting chest movement, with sensors placed on the body \cite{Al-Khalidi2011}. Breathing has been extracted using stretch sensors, an inertial measurement unit (IMU) strapped to the chest \cite{Reinvuo2006}, a head-mounted IMU + egocentric camera \cite{Hernandez2015}, and a wrist-mounted IMU \cite{Hernandez2015a,Sun2017}.




Non-contact breath measurement has been accomplished with a motion capture system \cite{Shafiq2017}, a smartphone microphone \cite{Ren2014}, and analysis of high-frequency wireless signals \cite{Ravichandran2015}. The latter approach enables multi-person monitoring, even with obstacles between people and transmitter \cite{Adib2015}.

The downside of remote methods is that they are constrained to a fixed environment. With Breeze we chose to use an IMU for portability and non-invasiveness.


\subsection{Digitally Presenting Biofeedback}
How to best display biofeedback is still an open question; no clear results exist in prior research on the tradeoffs between the various modalities, nor modalities' possible symbolic representations \cite{Chanel2015}.

Prior work investigates multi-modal biofeedback, as \cite{Wilson2016} with visual, audio, and haptic (vibrotactile and heat), as well as \cite{Roo2017} with visual, audio, and haptic (touch). In the latter work, touch is used as part of the interaction and not as feedback. Others describe shape-changing interfaces to represent breathing and heart rate \cite{Aslan2016}, as well as breathing robots \cite{Bucci2017, Yoshida2016}.

Biofeedback has been studied in social settings, such as sharing breathing to foster contentedness among couples \cite{Kim2015}, displaying cyclists’ heart rate on their helmet to support exertion \cite{Walmink2014}, and displaying the overall activation of the body on T-shirts \cite{Howell2016}. Biofeedback has also been used to help regulate anxiety by showing the link between biofeedback and a person's emotional state \cite{Costa:2016:EmotionCheck}.
Merrill et al. show that prior beliefs shape the comprehension of biofeedback \cite{Merrill2017}.



\pagebreak

\section{Breathing Encoded Across Multiple Modalities}
While prior work focuses on digitally presenting a user with their own breathing pattern, we aim to present breathing as biofeedback to others and have it understood emotionally. This section discusses how breathing can be digitally represented. 

Biofeedback can potentially be displayed in any modality (see Related Work) including taste and smell \cite{Ranasinghe:2011:DTS:2318776.2318795}. While not impossible to digitally create and reproduce, digital taste and smell are currently at early development stages, with issues around latency and inertia which make them unsuitable for real-time mapping of breathing.

This research was driven by the goal of designing the simplest representation within each modality. We map the entire range of breathing to one degree of freedom, thus supporting a more natural and intuitive interface. The following sections \linebreak discuss our design choices for mapping the normalized \linebreak (0 = no air in the chest, 1 = chest fully inflated) breathing signal from the IMU to the remaining candidate modalities: visual, audio, and haptic.

\subsection{Visual}
The simplest expression of a visual stimuli lies in control of a point light source, which can vary in color, brightness, and frequency (flicker). We discarded color because it might introduce a bias in the emotions' perception \cite{Stone2017}. Modulation of brightness of a point light seemed most adapted to visual biofeedback for breathing.

Since an increase in raw light intensity by a factor of two is not perceived by human senses as doubled brightness \cite{Poynton1993}, we applied a gamma correction, approximated by $brightness = breathing^{2.2}$, with both values normalized between 0 (minimum brightness = no air in the chest) and 1 (maximum brightness = chest fully inflated).

\subsection{Audio}
The most obvious audio feedback was to record a person’s breath, yet this would be complex to freely manipulate and render. Prior work used audio recording of the words ``breathe in'' and ``breathe out'' \cite{Yu2015a}, but these binary commands cannot grasp the variety of a continuous signal. Since our main requirement was to create a neutral sound, we had to avoid sounds with any connotation to objects, situations, and emotions. We chose pink noise as in \cite{Roo2017}.

To map breathing to \emph{loudness}, we took into account how sound is perceived. We used the OpenAL backend to convert the breathing to perceived loudness (in db) with $loudness = 10 \times log_{2}(breathing)$, and then drove the speaker with $gain = 10^{loudness/20}$.

\subsection{Haptic} 
Many different haptic sensations can be felt, including vibration \cite{Brewster:2004:TST:976310.976313}, tapping, rubbing \cite{Li:2008:TRE:1449715.1449744}, brushing \cite{Strasnick:2017:BEA:3025453.3025759}, and heat \cite{Wilson:2016:HUC:2858036.2858205}. While brushing, touch, and heat might become very natural communication channels for sharing biofeedback, they currently require bespoke hardware or present latency issues, making them unsuitable to convey continuous breathing data. As such, we decided to use vibration, a low-cost, widely available solution that is already used in the literature to support implicit communication between distant couples \cite{Bales2011}. We mapped the breath amplitude to the intensity (amplitude) of the vibration.


\begin{figure}
\centering
\includegraphics[width=1\hsize]{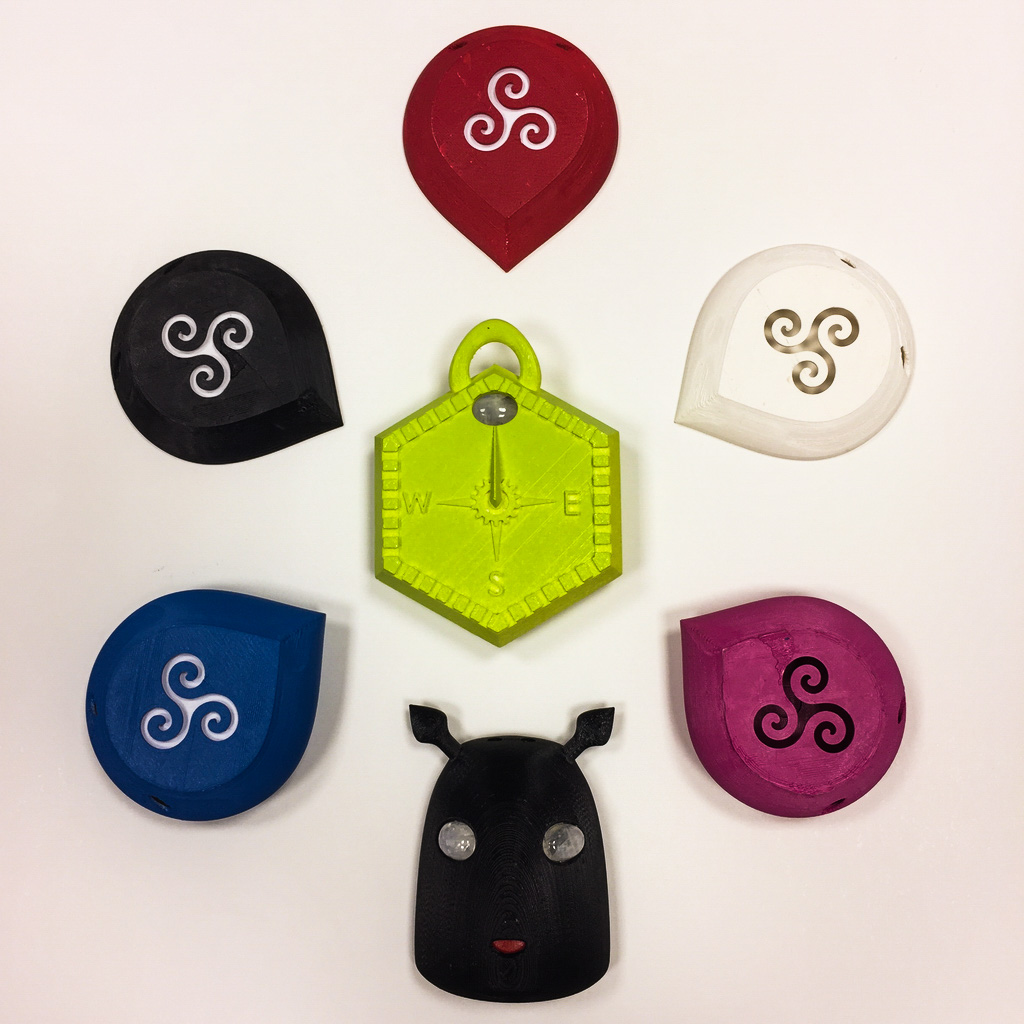}
\caption{Breeze 3D printed case designs: neutral teardrop shapes, a compass (center), and an avatar (bottom).}\label{fig:breezess}
\end{figure}

\section{Breeze: a Wearable Device for Breathing}
This section presents the design and development of Breeze (Figure \ref{fig:breezess}), a wearable device that monitors breathing in real-time and provides multi-modal biofeedback to the wearer. We describe the design process that led to the current form factor, the hardware embedded in the pendant, and the software used to get a live stream of the breathing pattern. Finally, we validate the pendant's breathing signal by comparing it with signals acquired through a more traditional breathing belt.

\subsection{Design}
In designing Breeze, we had to keep in mind the balance between capabilities and aesthetics. Breeze had to hold all electronics while remaining comfortable to wear. Our requirement was thus to measure breathing and provide feedback in a non-invasive, enclosed, and easily worn apparatus.

\subsubsection{Body Positioning}
Prior research shows that the position of the device on the body has an effect on the time required to access the interface \cite{Ashbrook:2008:QIM:1357054.1357092} and how the vibrations are felt in a mobile context \cite{Karuei:2011:DVA:1978942.1979426}.

With our requirements, only two body positions were suitable: the chest or wrist. We chose the chest as it provides better breath-sensing accuracy, given how much people move their wrists (e.g., typing, using a phone) during the day. Zeagler et al. \cite{Zeagler2017} indicated the upper part of the chest as a proper candidate to position wearables, and prior work proposes wearable pendants \cite{Amores2017a, Brueckner2014, Rennert2013} to act as either input or output.

\pagebreak

This choice created a few limitations:
\begin{itemize}
\item \textbf{Activity:} The sensing is not reliable while the wearer is walking. Since prior work shows that people are stationary around 70\% of the day \cite{Cauchard:2016:ADE:2858036.2858046}, we believe this limitation is acceptable.
\item \textbf{Field of View:} The light emitted is not in the user's active field of view. Yet, Harrison et al.'s work \cite{Harrison2009} shows that due to the amount of time spent working on a laptop and reading books, we are getting accustomed to looking down toward this region.
\end{itemize}

We envision that in a future version, users may sometimes untie the necklace to hold the pendant in their hands, or put it on their desk or bedside table as a companion. 


\subsubsection{Form Factor}
The form of the pendant affects both its perception and acceptability. We thought that for some an abstract design would work best, while others may prefer a symbolic representation or even an avatar that would be personified. We 3D printed several case designs (Figure \ref{fig:breezess}), including a neutral teardrop shape with the cutout ``air'' druid symbol, a compass with a glowing north, and an avatar\footnote{Inspired by Studio Ghibli's character Totoro} with glowing moonstone eyes. 

\subsubsection{Privacy Concerns}
Should biofeedback information be public or private? This question drove the design process of our system. Should Breeze be private or an artifact enabling social interactions? Our modalities cover both the private (haptic) and public spaces (visual and audio). In the current version, users can choose to remove the pendant at any time, mitigating privacy concerns from remote monitoring as in \cite{Sedenberg2017}. Future versions could allow users to select specific ``modes of interaction,'' choosing when to explicitly share biofeedback with others. 
Biofeedback data should be treated by default as private; pairing techniques between two Breeze pendants will need to be secure, and data encrypted in transmission. Users should also be made aware when their data is being collected/transmitted, e.g., through LED indicators.

%

\begin{figure}
\centering
\includegraphics[width=0.7\hsize]{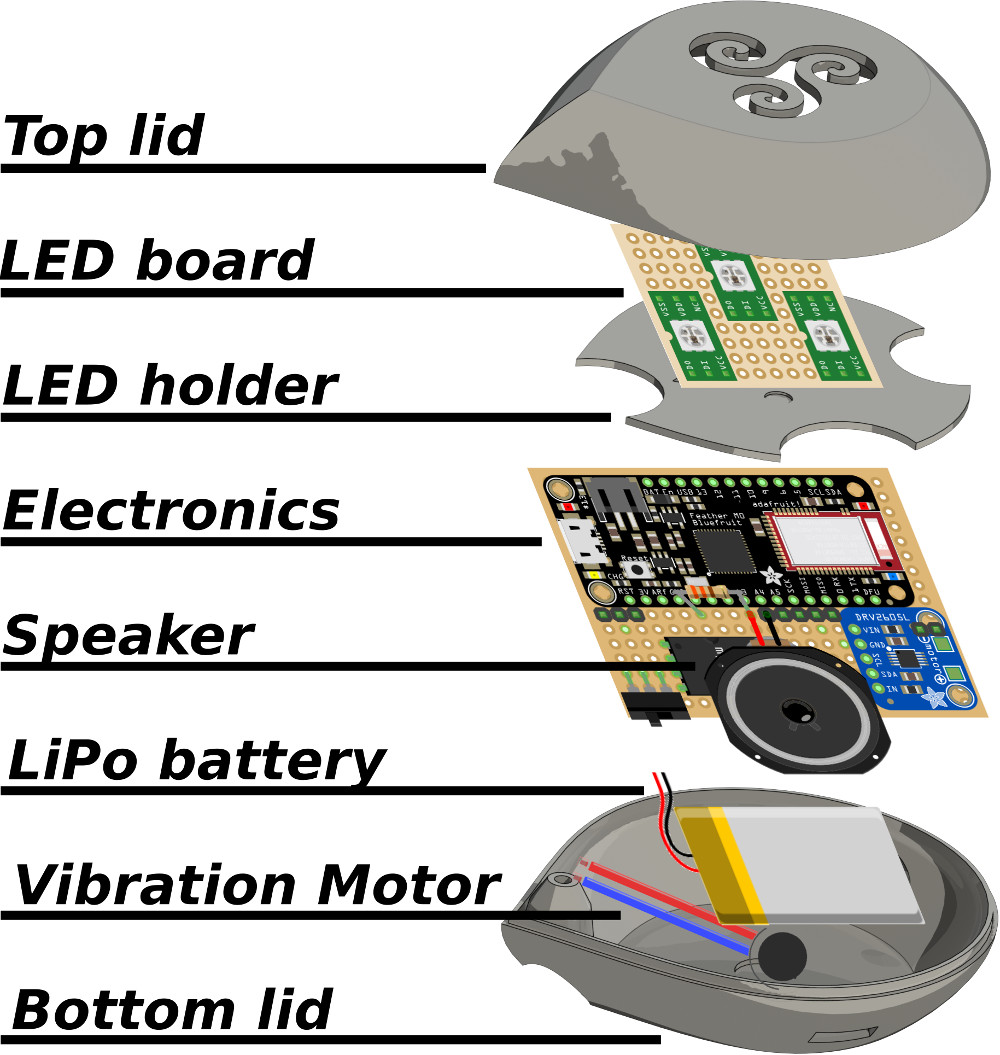}
\vspace{3mm}
\caption{Breakout of the components inside a Breeze pendant.}\label{fig:breeze-breakout}
\end{figure}

\subsection{Technical Implementation}
The pendants were 3D printed, and contained removable electronics consisting of an Adafruit Feather M0 BLE Arduino board, a 0.5W 8$\Omega$ speaker with 330$\Omega$ resistor, a vibrating motor controlled by an Adafruit DRV2605L haptic motor controller, 3 Adafruit Neopixel RGB LEDs, a 1000mAh LiPo battery for a full day of use, and a Sparkfun MPU-9250 9DOF IMU (Figure \ref{fig:breeze-breakout}). Each LED shone through a cutout with either a white diffuser or a moonstone. The vibration motor proved a design problem because it created noise within the pendant, which was not the intended modality; we eventually moved it to the outside back wall of the pendant. 

\subsubsection{On-board implementation}
We ran a sensor fusion algorithm \cite{Madgwick2010} on the Arduino in order to obtain absolute orientation and linear acceleration from the three IMU sensors: accelerometer, gyroscope, and magnetometer. Each signal alone is either noisy (magnetometer), drifts over time (gyroscope), or gives relative values (gyroscope and accelerometer). Yet, combined together, they provide accurate data in the terrestrial frame of reference.

\subsubsection{Real-Time processing}
We paired the pendant to a linux host computer running OpenViBE\footnote{\url{http://openvibe.inria.fr/}}. The sensor data (orientation and acceleration) was sent via Bluetooth at a sampling rate of 24Hz (due to the low bitrate of the BLE protocol) and the biofeedback returned at 10Hz. To extract breathing, we took the rotation in the pitch axis (pendant pointing toward the ground while on the chest) and processed the signal similarly to \cite{Hernandez2015a}: 0.5s time-based epoching, sliding window 0.125s, 1Hz Butterworth low-pass filtering (order 3), signal average of the 0.5s epoch.

\subsection{Measurement Validation}
To validate breathing measurements performed with Breeze, we compared the pendant side-by-side with a standard stretch-sensor breathing belt reproduced from \cite{Gervais2016}. During a twenty minute session, a participant wearing both the pendant and the breathing belt had to mimic onscreen breathing patterns in 30 trials of 40 seconds; we used the same patterns as described in the User Study section.

\begin{figure*}
\centering
\includegraphics[width=0.9\hsize]{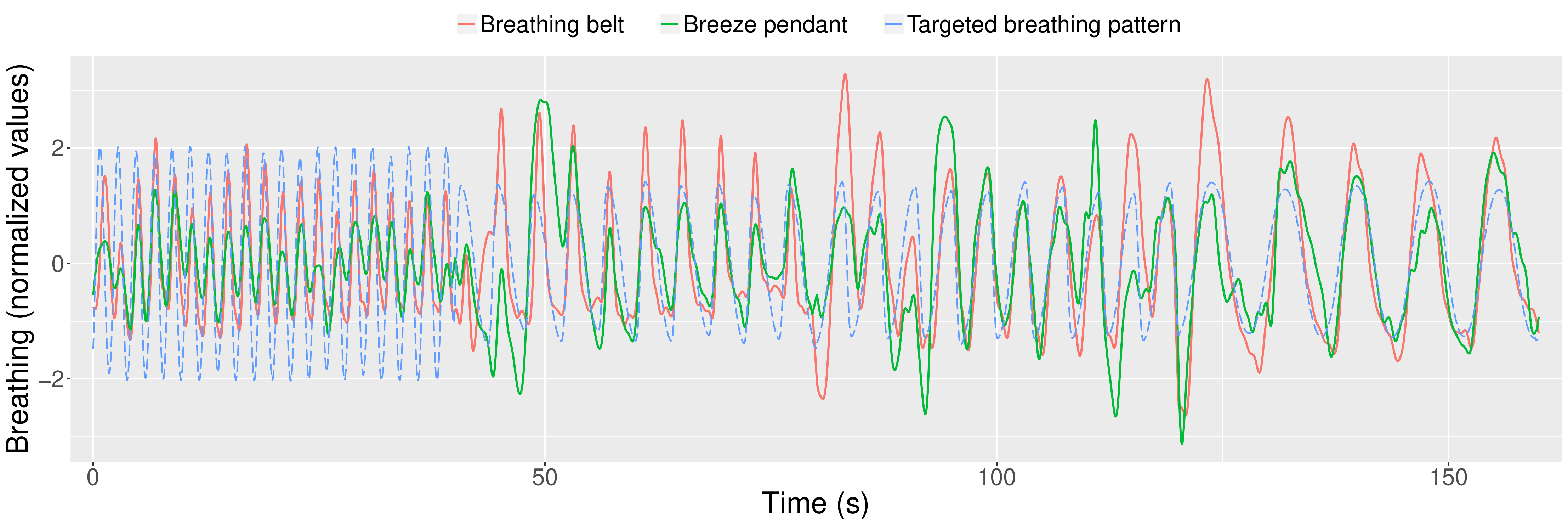}
\caption{Comparison between the Breeze pendant (green line) and a traditional breathing belt (red line). Over this 160s segment the participant was breathing at different speeds, from a faster pace to a slower pace. Both breathing measurements are correlated ($r = 0.54, p < 0.001$) and follow the patterns that the participant was mimicking (dashed blue line). (The values were normalized by standard deviation for visualization.)}\label{fig:validation}
\end{figure*}

Both signals were band-pass filtered between 0.1Hz and 1Hz and a 0.5s moving average was computed. Even using this straight-forward processing, signals matched, with a correlation $r = 0.54$ (Pearson correlation, p < 0.001) (Figure \ref{fig:validation}).

Differences between the two breathing signals are due to the nature of the sensors. The dynamic response to chest inflation of the conductive rubber stretch sensor in the breathing belt differs from an IMU's response. Rubber does not respond linearly to elongation and exhibits a ``memory effect'' when it recovers from being stretched (i.e., when the wearer starts to exhale). Despite those inherent differences, both breathing signals are correlated and display the same patterns.

\subsection{Fixed vs. Mobile Implementation} 
In order to study how well Breeze works in sharing breathing biofeedback, we first needed to understand users' perception of breathing across the different modalities. The results of the study will allow us to improve future versions of Breeze, which can then be deployed in longitudinal studies. In the design of the user study, we decided to use Breeze to monitor the breathing of participants and to provide vibrations. However, in order to better control the experiment, be consistent with the current literature, and ensure reproducibility of the results, we chose to study the audio and the visual feedback in a fixed setup, using earbuds and a monitor. This allowed us to study the emotional patterns without the influence of the hardware's novelty factor and its technical limitations.

\section{User Study}
This section details the user study conducted to determine how people perceive emotions conveyed by breathing patterns.

\subsubsection{Task}
\begin{figure}
\centering
\includegraphics[width=1\hsize]{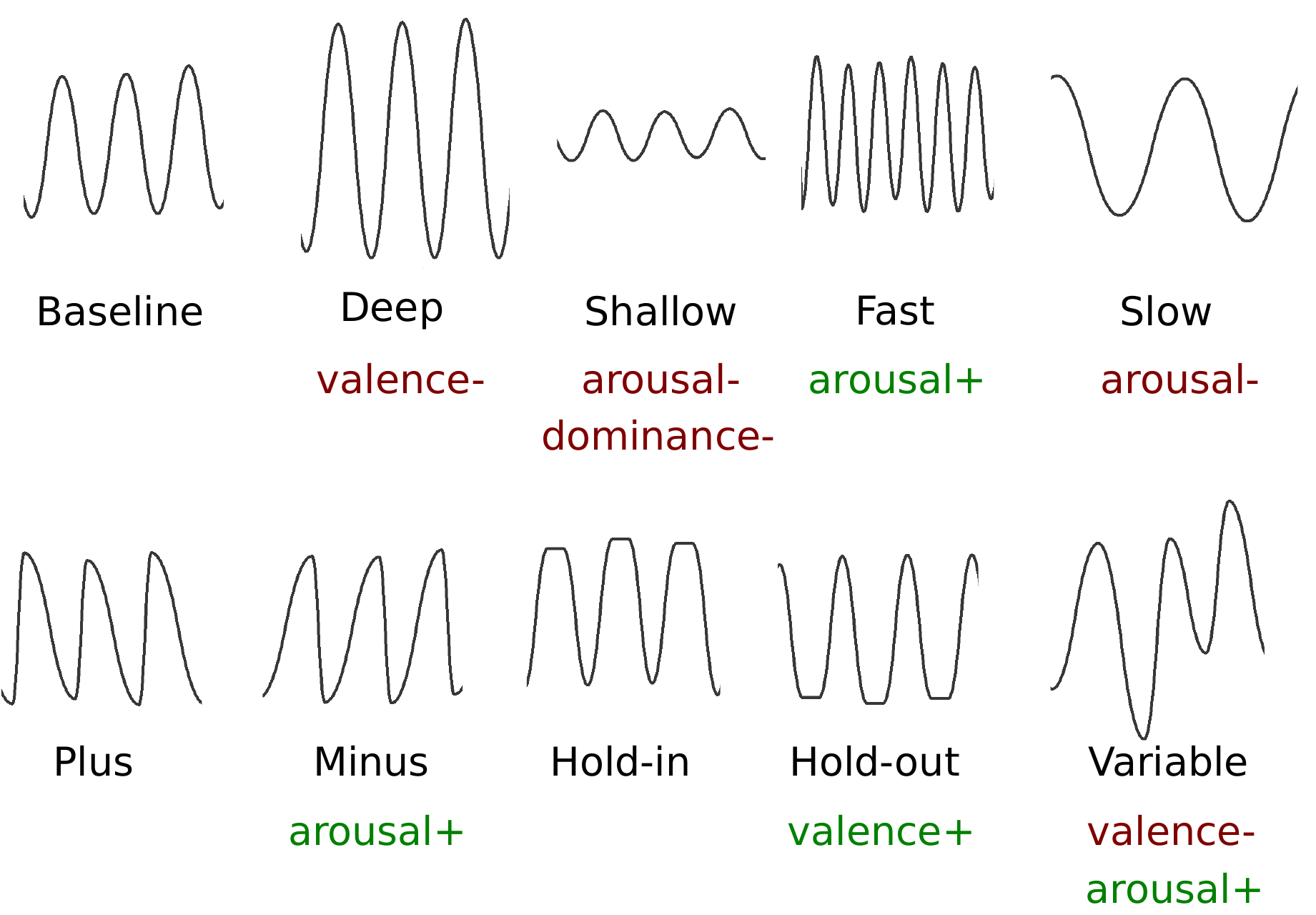}
\caption{Lexicon of breathing traits and corresponding emotional dimension analysis. The significant results from the SAM test \protect\cite{Bradley1994} are listed below each trait name, with the direction of significance indicated by a plus or minus sign.}\label{fig:lexicon}
\end{figure}

From our study of the literature, we selected the 5 \emph{features} defined below, whose variations are expressed as the 10 \emph{traits} detailed in Figure \ref{fig:lexicon}.

\begin{itemize}
\item \textbf{Pace of the breath.} Baseline: 15 resp/min; Fast: 30 resp/min; Slow: 7.5 resp/min.
\item \textbf{Difference between the time it takes to breath in and out.} Baseline: equal time; Plus: breath out takes 1s \linebreak longer than in; Minus: breath in takes 1s longer than out.
\item \textbf{Amount of time spent holding the breath.} Baseline: no holding; Hold-in: breath held for 2s after inhaling; \linebreak Hold-out: breath held for 2s after exhaling.
\item \textbf{Amplitude of the breath.} Baseline: 0.6; Deep: 1.0; \linebreak Shallow: 0.2 -- value normalized between 0 and 1.
\item \textbf{Variability of the pattern.} Baseline: no change; Variable: for each breath all parameters change randomly up to 50\% away from baseline.
\end{itemize}
To mitigate the task monotony, we generated breathing \emph{patterns} using combinations of these 10 traits and presented them to the participants. For instance, Hold-in and Hold-out were combined with Slow, and Deep and Shallow with Fast, in a total of 10 patterns (see final list of patterns in Figure \ref{fig:sam-patterns}).


After each trial, participants answered a short questionnaire to assess the emotion of the person that was (supposedly) breathing. Using the Self-Assessment Manikin (SAM) test \cite{Bradley1994}, they rated the three emotional dimensions: valence, arousal, and dominance. Each dimension was measured on a 9-point Likert scale represented by the SAM characters, as choosing between pictures instead of using words helps people to express feelings that could be difficult to externalize.
\emph{Valence} relates to the hedonic tone and varies from negative to positive emotions (e.g., frustration \emph{vs} pleasantness);
\emph{arousal} relates to bodily and mental activation and varies from ``calm'' to ``excited'' (e.g., satisfaction \emph{vs} happiness);
\emph{dominance} relates to the degree of control and varies from ``submissive'' to ``in control'' (e.g., afraid \emph{vs} angry). 

\subsubsection{Breathing analysis}
The breathing of the participants was covertly monitored with the pendant, as it was in constant contact with the chest during the experiment. The signals were streamed to the computer and synchronized with the beginning and end of each 40s trial using the lab streaming layer (LSL)  protocol\footnote{\url{https://github.com/sccn/labstreaminglayer}}.

Afterwards, we processed the participants' data to extract the peaks, from which we calculated the following breathing signal features: pace of the breath, difference between the time it takes to breath in and out, amount of time spent holding the breath, and amplitude of the breath). To the rotation of the pitch axis, we applied a 0.1Hz Butterworth high-pass filter (order 3) to remove the offset while preserving the signal's shape, applied a moving average, and used a first order derivative to detect the peaks within the smoothed signal.

\subsubsection{Post-study questionnaires}
After the completion of the main task, participants were given a NASA-TLX questionnaire \cite{Hart1988} to measure the mental, physical, and temporal demand; performance; effort; and frustration. Summing the various answers results in a task load index normalized between 0 and 100. 

Two additional questionnaires inquired about how much they enjoyed each modality, and how useful the modality was in assessing emotions (5-point Likert scales with only the extremes labeled, as ``Not at all'' and ``Extremely'').

\subsection{Protocol}
Fourteen participants (7f, 6m, 1 non-binary, Mean = 25 y.o. (SD: 3.57)) took part in the study. Participants were welcomed into the experiment room and signed a consent form.


The experiment was divided into a training phase and an experiment phase. During the training phase ($\approx$ 5 minutes) participants were explained the function of the pendant, shown the questionnaire, and given sample trials using all three biofeedback modalities (visual from the computer screen; audio using ear-buds; and haptic by wearing Breeze with noise-canceling ear muffs so as not to be distracted by the sound caused by the vibrating motor). The mapping from chest inflation to each biofeedback modality was also explained.

During the experiment phase, each participant experienced, in random order, 30 trials (10 patterns $\times$ 3 modalities) each lasting 40 seconds.

Afterwards, participants answered the post-study questionnaires and participated in a semi-structured interview. The experiment lasted 45 minutes on average. Participants were compensated 10 USD (in local currency) for their time.

\begin{figure*}
\centering
\subfloat[\label{fig:sam-patterns}]{\includegraphics[width=0.49\hsize]{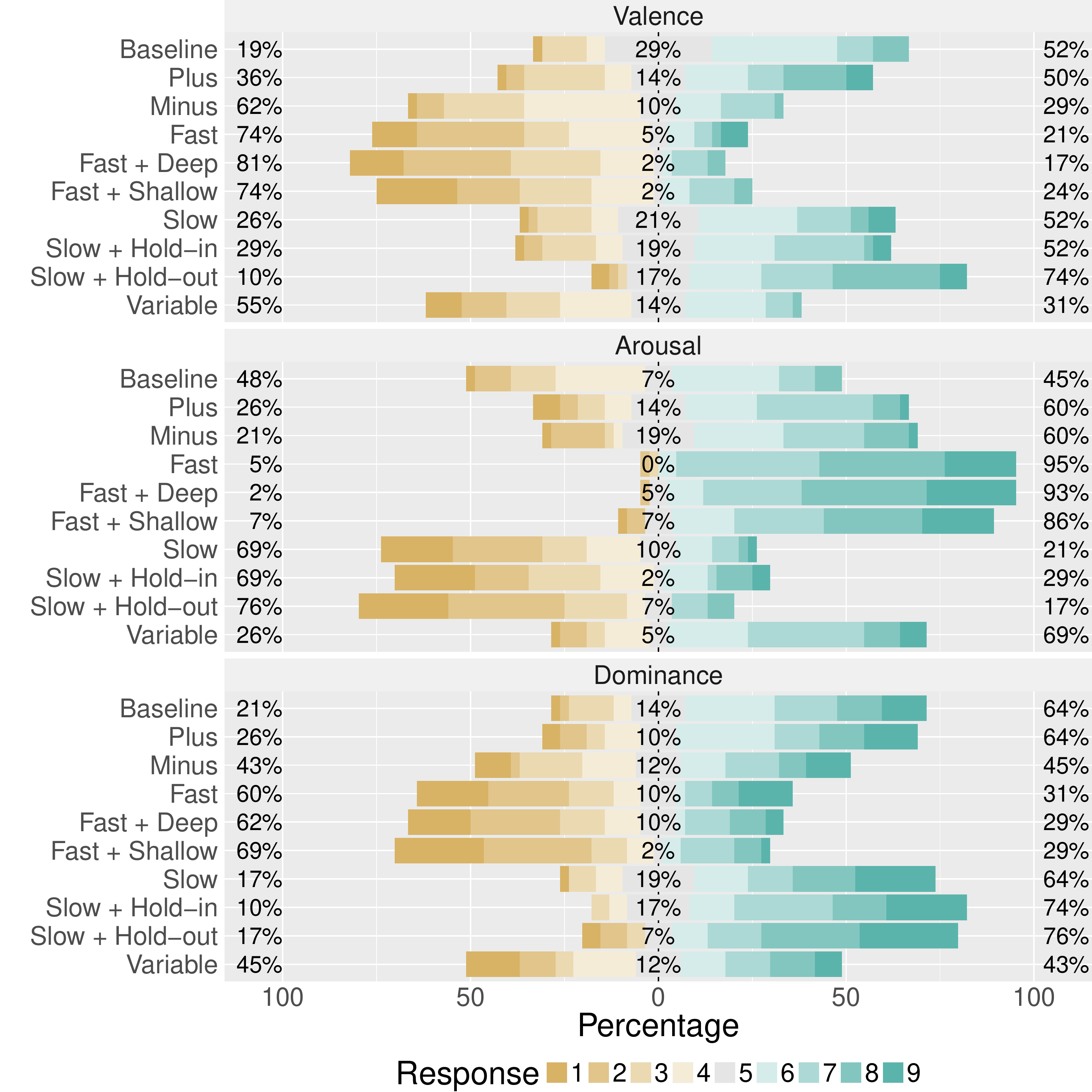}}
\hfill
\subfloat[\label{fig:sam-feedback}]{\includegraphics[width=0.49\hsize]{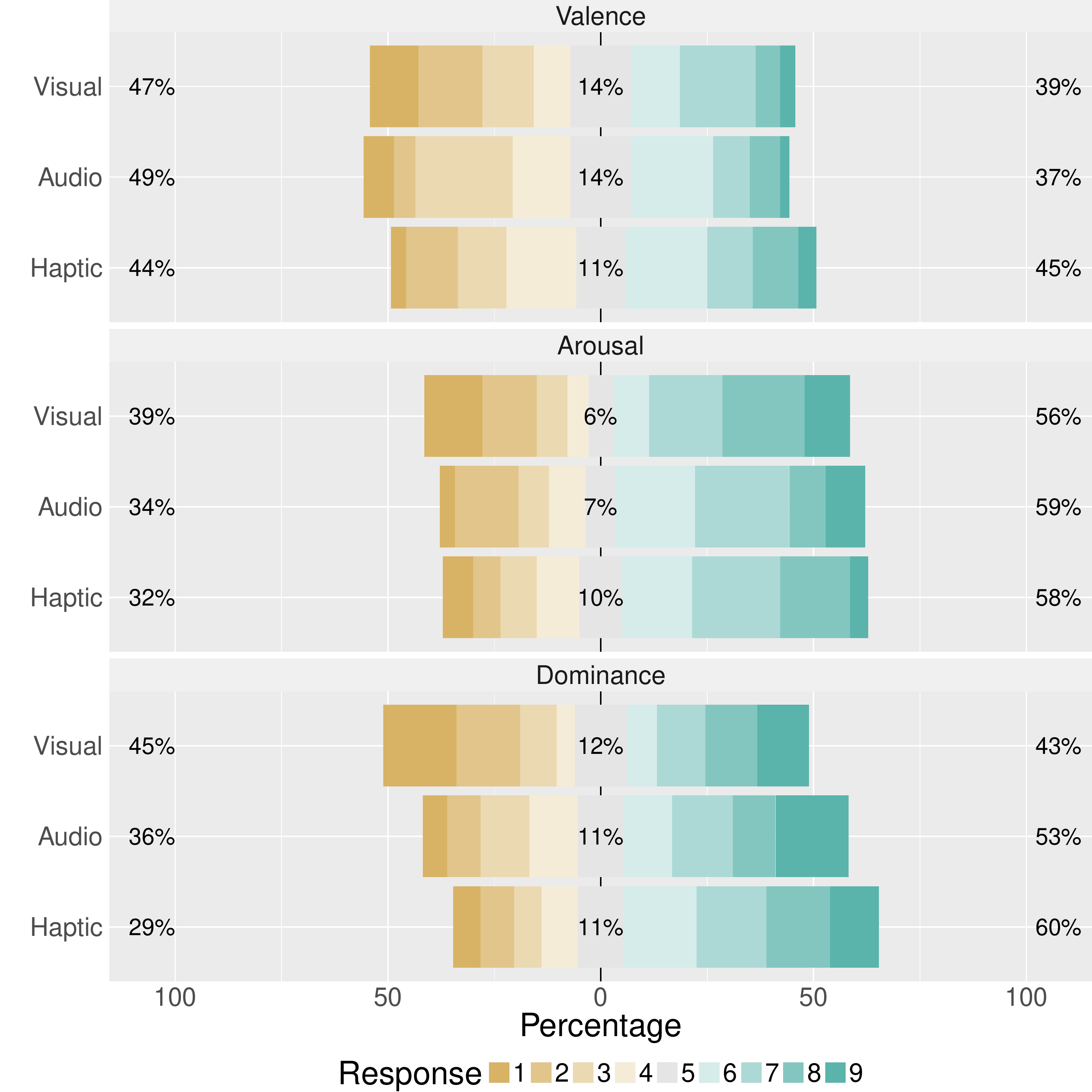}}
\caption{SAM questionnaires \protect\cite{Bradley1994}: participants' emotional dimensions' ratings for the ten generated breathing patterns (\emph{\ref{fig:sam-patterns}}) and feedback modalities (\emph{\ref{fig:sam-feedback}}). Proportions are given below midpoint, at the midpoint of 5, and above midpoint.}\label{fig:sam}
\end{figure*}

\subsection{Results}
The results are presented across four sections: analysis of the perceived emotions, analysis of the breathing, analysis of the post-study questionnaires, and finally reports from the semi-structured interviews.

\subsubsection{Effects of breathing traits and feedback modality on emotion}

Our analysis measured the effect of breathing traits \linebreak (Figure \ref{fig:sam-patterns}) and feedback modality (Figure \ref{fig:sam-feedback}) on emotion. 

We used a Markov Chain Monte Carlo (MCMC) method \cite{Hadfield2010b} which included SAM emotional dimensions as response (dependent variables); breathing traits and feedback modalities as fixed effects (independent variables); participants as random effect; and no intercept. It was parametrized using gaussian distributions, with both variance and covariance computed for residuals and variance for random effects. Due to the stochastic nature of MCMC we controlled for the results' convergence by using Gelman and Rubin's Convergence Diagnostic \cite{Brooks1998} on 10 chains. The resulting multivariate potential scale reduction factor (MPSRF) was 1.002.

All emotional dimensions measured with the SAM questionnaires were significantly (p < 0.05) affected by at least one breathing trait. The breathing traits (Figure \ref{fig:lexicon}) were compared to the Baseline pattern:

\begin{itemize}
\item \textbf{Valence:} Affected \emph{positively} by longer amount of time spent exhaled (Hold-out: mean = +1.43, 95\% CI [0.35, 2.47]); \emph{negatively}  by a more variable breathing (Variable: mean = -1.16, 95\% CI [-2.20, -0.09]) and by a deeper amplitude (Deep: mean = -1.31, 95\% CI [-2.39, -0.27]).
\item \textbf{Arousal:} Affected \emph{positively} by a higher pace (Fast: mean = +3.33, 95\% CI [2.25, 4.38]), by more variable breathing (Variable: mean = +1.11, 95\% CI [0.02, 2.17]), and by a lower ratio between exhalation and inhalation (Minus: mean = +1.10, 95\% CI [0.02, 2.15]); \emph{negatively} by a slower pace (Slow: mean = -1.24, 95\% CI [-2.29, -0.15]) and by a shallower amplitude (Shallow: mean = -1.08, 95\% CI [-2.13, -0.02]).
\item \textbf{Dominance:} Affected \emph{negatively} by a shallower amplitude (Shallow: mean = -1.25, 95\% CI [-2.42, -0.08]).
\end{itemize}

In Figure \ref{fig:sam-patterns}, we see that breathing pace (Fast and Slow traits) seems to have the biggest influence among breathing features, probably because it was the easiest change in pattern to notice and as it was interleaved with others. Yet, the effect of pace on perceived valence and dominance was marginal (p < 0.01) and only its effect on arousal was significant (see above).

We found a significant effect of the haptic modality on the perceived emotions, which elicited a higher valence overall (mean = +0.88, 95\% CI [0.03, 1.77]). However, we did not observe a significant effect of the audio and visual modalities on the perceived emotions.

Preliminary findings indicate that the SAM rating for Deep differs significantly with audio compared to other modalities, and for Fast with haptic. A larger study would help confirm these results. Due to the small sample size, we could not fully investigate the interactions between each output modality, each breathing trait, and each emotional dimension.

\subsubsection{Change in breathing}

To analyze how the generated patterns presented to participants affected their breathing, we compared the breathing signal features measured through the pendant with the breathing traits. The MCMC analysis included participants' breathing signal features as response (dependent variables); breathing traits as fixed effects (independent variables); participants as random effect; and no intercept. \linebreak Results converged, MPSRF = 1.002.

We observed a significant effect (p < 0.05) of both pace as well as the ratio between inhalation / exhalation (i.e., amount of time spent holding the breath). The breathing rate across participants was 16.35 breaths/minute on average (95\% CI [14.81, 17.93]), on par with the average for their age group \cite{Gomez2016}, and on par with the Baseline pattern, which was set to 15 breaths/minute. When observing a pattern with the Fast trait, the breathing rate of participants increased +2.04 breaths/minute (95\% CI [0.63, 3.42]), and decreased -2.64 (95\% CI [-4.04, -1.28]) with the Slow trait. It should be noted that when observing a Variable pattern (i.e., signal with a higher variability), the participants' breathing rate also decreased, by -1.44 (95\% CI [-2.85, -0.06]).

In patterns composed of the Minus trait, when the ratio between the inhalation and the exhalation was lower, participants followed the pattern and spent slightly more time exhaling than inhaling (+2.05 seconds, 95\% CI [0.48, 3.59]).

\subsubsection{Post-study questionnaires}
While all fourteen participants were interviewed, due to adjustments in the protocol, only twelve answered the NASA-TLX questionnaire.

The average task load index exacted from the NASA-TLX was 34.2\% (SD: 17.7), indicating a low to moderate effort \cite{Grier2015}.

\begin{figure}
\centering
\includegraphics[width=1\hsize]{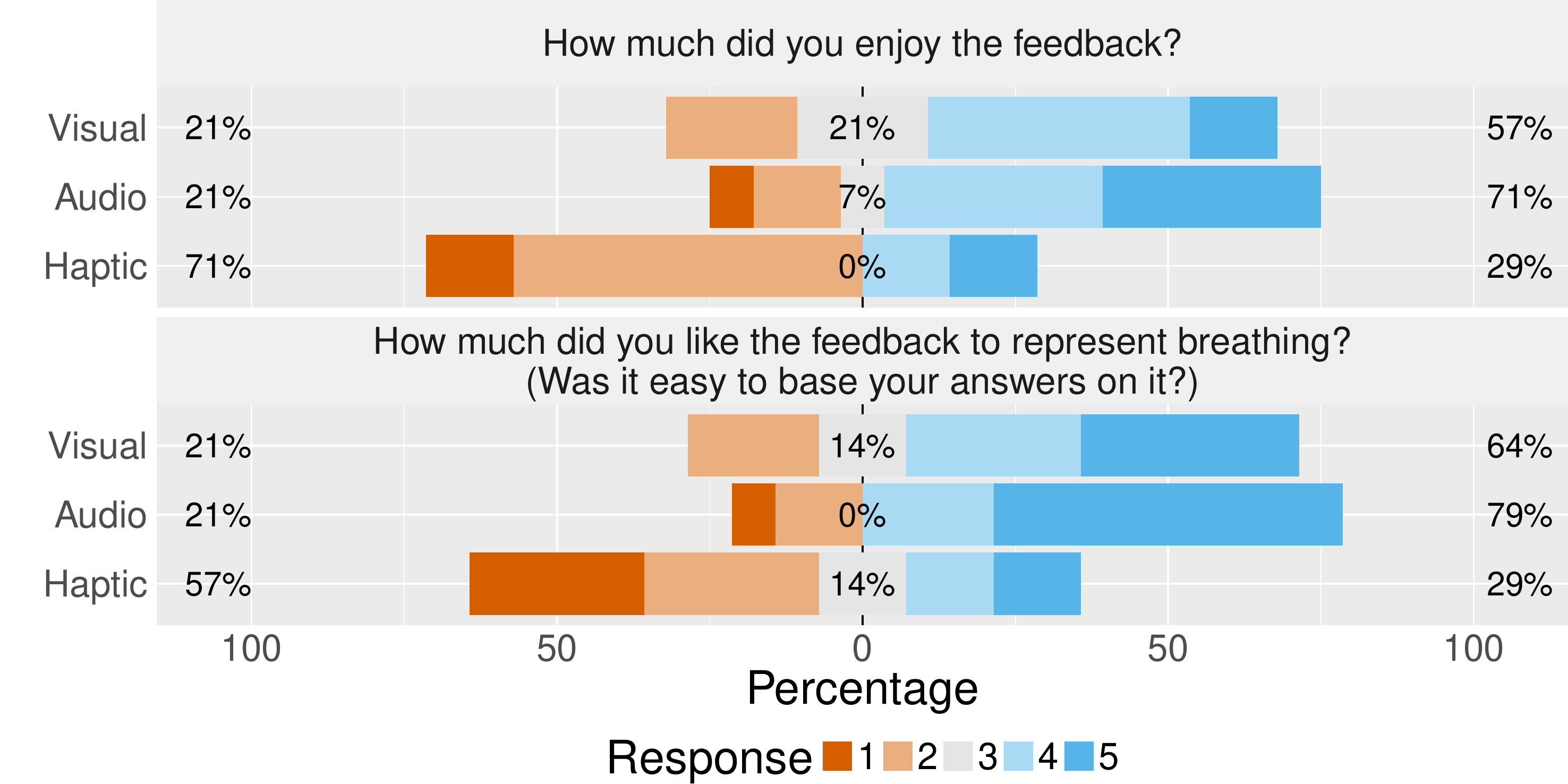}
\caption{Post-study questionnaires. Proportions on the left are below midpoint, in the middle at the midpoint of 3, and on the right above midpoint.}\label{fig:likerts}
\end{figure}

We used an MCMC method analogous to the above to analyze the 5-point Likert scale questionnaires. Haptic feedback was rated significantly lower than audio and visual feedback on both enjoyability and usefulness (p < 0.05, see Figure \ref{fig:likerts}), scoring $-1.22$ (95\% CI [-2.18, -0.29]) on enjoyability as compared to an average of $3.80$ (95\% CI [3.11, 4.45]), and scoring $-1.51$ (95\% CI [-2.55, -0.47]) on usefulness as compared to an average of $4.08$ (95\% CI [3.33, 4.79]).

\subsubsection{Interviews}
Next we describe the semi-structured interviews' results.

\par{\emph{Feedback modalities}}
Participants were asked to describe their experience for each modality.

\pagebreak

\begin{itemize}
\item \textbf{Visual:} A majority of participants expressed positive feedback and described it being “easier to imitate” (i.e., mimic). One person explained that it challenged their concentration abilities and was more difficult to follow than the audio modality which was clearer. 
\item \textbf{Audio:} A majority of participants expressed positive feedback, referring to the sound as being “relaxing” and associating the modality with the sound of waves. They found this modality “easy” to understand and mimic. Two participants disliked the modality, which felt like “noise” and was difficult to understand.
\item \textbf{Haptic:} Nine participants expressed negative feedback, three linking the vibrations to phone notifications. Two were neutral; one described vibrations as being ``very calming'', another stated that they ``could relate better''.
\end{itemize}

\par{\emph{Breathing Pace and Perception}}
All but one participant referred to determining the individual’s level of arousal and dominance based on the breathing pace. Several noted that a faster pace indicated higher arousal but lower dominance, and similarly that a slower pace indicated higher dominance. While some participants associated faster breathing with stress, others associated it with excitement. Slower breathing patterns were associated with relaxation by several participants, taking into account both the pace and the breathing pattern itself. One person explained that if the pattern started with a high pace and slowed down, the individual was not in control, whereas if the already fast-paced pattern increased, the emotion they felt simultaneously intensified. 

\par{\emph{Breathing Context}}
We asked participants if they could identify the context in which particular breathing patterns were recorded. With shallow breathing, half of the participants mentioned someone exercising or running, whereas the other half associated it with a stressed or frightened state. With slow breathing, they associated the pattern with sleep, relaxation, meditation, and tranquility. Four participants said that they did not think about any particular context during the experiment.

\par{\emph{Mimicking}} 
All but one participant explained actively trying to mimic the breathing pattern in order to understand it. One mentioned that breathing at the same pace helped understand the context. Others said that they would not be able to determine what ``the other person'' (the generated breathing pattern) was feeling without mimicking.

\par{\emph{Usages}}
In the final stage of the interview, we introduced the notion of using Breeze as a way to maintain a bond with a significant other, each one perceiving the other's breathing.

\begin{itemize}
\item \textbf{Why?} Participants envisioned using Breeze in long distance relationships with people close to them, such as a partner or family member. One participant noted that Breeze was ``too intense'' for usage with friends, even ones they identified as being close to them, due to the intimacy factor. 

\item \textbf{Who?} Participants envisioned using Breeze with a life partner (5), a close family member (5), and more specifically a child (2). One participant mentioned their grandmother who is currently going through cancer treatment as it would be ``nice to know how she feels.'' Only one participant did not have anyone with whom they would like to share breathing biofeedback. 

\item \textbf{What?} Most participants said they would use Breeze to convey their current emotional states, with two participants mentioning that they would use it to detect a negative change in their loved one's mood, to be able to positively influence them. One participant explained that this communication channel is more intimate than texting.

Almost half of the participants envisioned emergency-related situations such as health monitoring. Participants also mentioned they could use Breeze to: convey feelings that are not easily expressed; show people how their actions can affect those around them, including bullying; reflect on their own biofeedback; and prevent negative states.
\end{itemize}

\par{\emph{{Privacy}} Four participants said that they would not use this device because of privacy issues, with one person not understanding the point of connecting people using technology. Yet, most participants (12) mentioned that they would be comfortable sharing their breathing activity or heart rate in this manner, one of them comparing it to sharing status on social media platforms.
 
\par{\emph{Case Design}} Almost half of the participants preferred a simple colored pendant for daily wear and several liked the idea of a compass as finding one's way in life. The avatar design was generally seen as ``cute'' and more suitable for children, while two participants referred to it as ``frightening''. Most participants found the pendants ``too large'' and people not used to wearing jewelry preferred the idea of a small pendant that can be tucked into their shirt. Some participants envisioned a shape that would represent their loved ones, with two mentions of a red heart-shaped pendant.

\subsection{Discussion}
This section presents a discussion around the quantitative and qualitative results of the study. 

\subsubsection{Breathing to communicate}
We hypothesized that exposing participants to breathing biofeedback would affect their own breathing. This assumption held true, as the participants' breathing rate changed to match the pace of the displayed patterns, without any instruction. Almost all participants reported consciously mimicking the biofeedback in order to better comprehend what ``the other person'' was experiencing. The exception was a participant who started by breathing simultaneously with the pattern and then stopped because they thought that the purpose of the experiment was to try ``to control [their] breathing.'' We were surprised by how empathetic participants became toward simple unimodal and unidirectional cues, treating simple stimuli as another person's biofeedback.

This mimicking action provides interesting insights in alternative ways to communicate at a distance. Should the activity of the inner body be made explicit, people might naturally use it to understand others. For instance, Philippot et al. \cite{Philippot2002} \linebreak describe similar breathing between acting an emotion and feeling one. Hence to reproduce emotions might well be a coherent strategy to assess others' states.

\subsubsection{Breathing features}
Prior work in this field focused on breathing pace and overall breathing variability. We found that the advanced features we extracted from participants wearing Breeze were meaningful: for example, the time spent breathing out matched the source biofeedback during mimicry. We hope to raise the interest of researchers about the amount of data that lies within breathing. Our lexicon (Figure \ref{fig:lexicon}) could be used as a starting point for extracting  meaningful information from the breathing signal.

\subsubsection{Biofeedback modalities}
The pink noise chosen in the audio modality was often interpreted as wave sounds, a phenomenon associated with a calming effect. This association, however, did not significantly affect perception of the patterns.

While several participants reported that they could not interpret the breathing pattern as efficiently via the haptic modality, no such effect was seen on the SAM test. The discrepancy between perceived and actual effectiveness might be due to the fact that this modality is not traditionally used to convey meaningful information. Similarly, while the vibrations led to negative connotations for some participants, this did not negatively bias the emotion perceived. Instead, the breathing patterns perceived through haptic feedback translated into a slightly higher valence.

\section{Challenges and Limitations}

While we find some differences between the three modalities, a larger sample would be needed to investigate whether they are equally meaningful and effective in conveying biofeedback.

In the future and in order to deploy this work for a longitudinal study, we will need to reduce the size of Breeze using a custom PCB, and port the signal processing to a mobile device. We could explore different form factors, such as a smartwatch. The peak detection will need to be made more robust to movement. Although considering the sedentary Western lifestyle \cite{Cauchard:2016:ADE:2858036.2858046}, Breeze's current algorithm should be mostly sufficient, especially since the pendant stays in contact with the chest even when users are leaning slightly forward.

Future work might consider extending to measure both thoracic and abdominal respiration -- preliminary observations show that this data is encoded within the current IMU's signal, and thus could be used to add more information about the breathing pattern. One could also extract heart rate from the acceleration as per \cite{Hernandez2015,Hernandez2015a}.

Even though it did not affect how people perceived emotions, participants' comments about haptic feedback should be taken into consideration. As vibrations will keep being used for pushing - disrupting - notifications, haptic feedback should be revised to appear more neutral. An alternative form of actuation, such as inflation \cite{Bucci2017}, could be used, with the added benefit of being noiseless.

Now that the protocol is established, the study of the perceived emotions could be extended with a wider repertoire of breathing patterns. Finally, we could extend the protocol to go beyond emotions and toward metaphors \cite{Seifi2015}, or match richer adjectives with the patterns \cite{Bucci2017}.

\section{Future Work and Conclusions}
We described Breeze, a wearable device to communicate breathing biofeedback. Breeze functions bidirectionally, by collecting data with physiological sensors and providing ambient biofeedback. 

To assess what information people can infer from the various feedback modalities that Breeze provides (visual, audio, and haptic), we conducted a laboratory study where people rated the perceived emotions from a set of generated breathing patterns. To our knowledge our work is the first to describe, quantitatively (breathing measurement) and qualitatively (semi-structured interviews), how people naturally mimic a foreign breathing pattern in order to understand it. We described a simple yet effective methodology for extracting interactions between breathing and perceived emotions, opening the use of breathing as a form of biofeedback.

Such findings may reinforce the bond that shared biofeedback can provide in remote communication. We envision that making physiological signals more visible could promote empathy and improve connectedness. Our next step is to deploy this technology outside the laboratory in a long-term longitudinal study. We look forward to seeing how people will use Breeze in their everyday lives to communicate with their loved ones.

\subsection{Acknowledgments}
The authors would like to thank Esther Mandelblum for her work on the visual material, as well as Andrey Grishko for his early designs of the Breeze cases. We would also like to thank the reviewers for their time and insightful comments.

\balance{}

\bibliographystyle{SIGCHI-Reference-Format}
\bibliography{sample}

\end{document}